\documentclass[fleqn,usenatbib]{mnras}

\usepackage{newtxtext,newtxmath}

\usepackage[T1]{fontenc}
\usepackage{ae,aecompl}

\usepackage{graphicx}	
\usepackage{amsmath}	
\usepackage{amssymb}	

\title[Modelling the MIR polarization around young stars]{Modelling the mid-infrared polarization in dust around young stars}

\author[G. Saikia et al.]{
Gautam Saikia,$^{1}$\thanks{E-mail: gautamsaikia91@gmail.com}
Rupjyoti Gogoi,$^{1}$\thanks{E-mail: rupjyotigogoi@gmail.com}
Ranjan Gupta$^{2}$
and Deepak B. Vaidya$^{3}$
\\
$^{1}$Department of Physics, Tezpur University, Napaam-784028, India\\
$^{2}$IUCAA, Post Bag 4, Ganeshkhind, Pune-411007, India\\
$^{3}$Ex-Gujarat College, Ahmedabad-380006, India
}

\date{Accepted 2019 January 18. Received 2019 January 18; in original form 2018 August 15}

\pubyear{2019}

\begin{document}
\label{firstpage}
\pagerange{\pageref{firstpage}--\pageref{lastpage}}
\maketitle

\begin{abstract}
The presence of crystalline silicates has been detected in the circumstellar environment of several young stars in the recent past and there is evidence of silicon carbide (SiC) detection in the envelope of pre-main sequence star SVS13. In this work, we have attempted to probe the presence of SiC in the dust around protoplanetary disks in a sample of young stars. We have modelled the linear polarization of composite dust grains in the mid-infrared (MIR: 8--13 $\mu$m) using silicates as the host with various inclusions of SiC and graphites using the Discrete Dipole Approximation (DDA) and the Effective Medium Approximation (EMA) T-Matrix methods. We have then compared our modelling results with polarimetric observations made in the protoplanetary disks surrounding two Herbig Be stars and one T-Tauri star with particular emphasis towards the 10 $\mu$m silicate feature using CanariCam mounted over the Gran Telescopio Canarias (GTC). We report the possible existence of SiC in the outer disk/envelope around one star in our sample which has been interpreted based on the shape, size, composition and fraction of inclusions by volume in our dust grain models.
\end{abstract}




\section{Introduction}

Herbig Ae/Be stars, first identified by \citet{Herbig+1960}, are young (1--10 Myr) objects in the pre-main sequence having masses between 2--10 M$_\odot$. Herbig Ae/Be stars are usually enclosed by a disk and/or a gaseous, dusty envelope since they form a part of the process of star formation [\citet{Waters+Waelkens1998}]. Due to the infrared excess arising from the disk/envelope of Herbig stars [\citet{Waters+Waelkens1998}; \citet{Leinert+2001}], they are often regarded as the massive analogues to young lower mass T-Tauri stars [\citet{Juhasz+2010}]. \citet{Meeus+2001} suggested that circumstellar disks surrounding Herbig stars could be categorized into two main groups: Group I comprising of sources showing strong infrared excesses having flared disks (mostly Herbig Ae stars) and Group II comprising of sources that show modest infrared excesses having flatter, self-shadowed disks (mostly Herbig Be stars) as confirmed first by \citet{Leinert+2004} and \citet{Acke+2005}. The disk properties in Group I Herbig Ae stars have been observed to be very similar to T-Tauri stars [\citet{Boekel+2008}; \citet{Alonso-Albi+2009}]. The study of protoplanetary dust in these stars gives us an opportunity to understand massive star formation as well as planet formation processes [\citet{Alonso-Albi+2009}] and the observation of large infrared excess makes MIR polarimetry an ideal tool to study their dust grain properties. Polarization is represented as the difference between the parallel and perpendicular components of extinction efficiency which are independently related to the dust optical constants [\citet{Whittet+2003}]. The polarization profile (peak wavelength, shape and strength) is highly susceptible to the specific shape, composition as well as dust grain size and hence, the study of polarization across an absorption/emission feature can be very efficiently used to constrain these properties [\citet{Dullemond+2008}; \citet{Fujiyoshi+2015}; \citet{Wright+2016}].\\

The composition of dust particles found in the protoplanetary disks around young stars can be reasonably assumed to be analogous to those in the interstellar medium (ISM) including silicates which are the most abundant species found in the ISM [\citet{Min+2007}; \citet{Molster+2010}]. The peak in absorption which arises due to Si--O bond stretching near MIR 10 $\mu$m has been widely studied and this feature, visible clearly in the linear polarization profile of interstellar dust grains, is attributed to non-spherical and aligned silicate dust grains [\citet{Whittet+2003}; \citet{Min+2007}]. While amorphous silicates present in the ISM are mostly responsible for the 10 $\mu$m peak, \citet{Aitken+1988} observed an additional feature around 11 $\mu$m in the polarization profile towards a Class I Young Stellar Object (YSO), AFGL 2591, which was considered to have formed due to crystalline silicates similar to olivine. Separate studies have placed an upper limit on the abundance of silicates found in crystalline form in the ISM: 2.2\% by \citet{Kemper+2004,Kemper+2005} and 1\% by \citet{Min+2007}. More recently, \citet{Wright+2016} have detected a 2.5--5\% crystalline abundance towards the Galactic centre based on three features at 11.1, 11.9 and 23.5 $\mu$m, which is consistent with the upper limit of \citet{Kemper+2005} and within the 3--5\% limit set by \citet{Li+2007}. In case of protoplanetary disks, a review by \citet{Henning+2010} has stated that amorphous silicates with various compositions are expected to be found in the outer regions while crystalline dust is expected towards the innermost (high temperature $\sim$ 1300 K) regions of the disk with the existence of annealed amorphous grains in-between the two regions. But the fact that crystalline material has also been observed in low temperature ($\sim$100 K) outer regions of disks means that there must either be some mechanism for large-scale transport/mixing or for low-temperature crystallization [\citet{Juhasz+2010}; \cite{Shu+2011}].\\

The existence of crystalline silicates has now been observed in protoplanetary environments of both  T-Tauri and Herbig Ae/Be stars [\citet{Kessler-Silacci+2006}; \citet{Sicilia-Aguilar+2007}; \citet{Bouman+2008}; \citet{Meeus+2009}; \citet{Olofson+2009}; \citet{Watson+2009}; \citet{Juhasz+2010}]. The interest in study of silicon carbide (SiC) as a potential dust constituent in protoplanetary disks arises from the fact that SiC, which is known to be formed in the circumstellar environment of AGB stars rich in carbon, has been found in pre-solar samples, i.e. all SiC grains are not destroyed in the ISM [\citet{Bernatowicz+2006}]. Hence, pre-solar SiC serves as an ideal tool to study new planetary systems and how they were formed with the same dust as that produced in aging stars [\citet{Hofmeister+2009}]. SiC has also been found as a dust constituent in comets and meteorites with evidence of pre-solar SiC grains to have been formed around the same time as graphites in the sample or even earlier [\citet{Orofino+1994}; \citet{Messenger+2009}; \citet{Floss+2013}]. On the basis of a feature occurring around 11 $\mu$m, \citet{Min+2007} found $\sim$3\% SiC in interstellar dust grains with 9--12\% of the silicon present in the SiC, but their inference was purely model-dependent. Several candidates have been identified for the 11 $\mu$m feature including SiC, hydrocarbons, water ice, carbonates and crystalline silicates. \citet{Pott+2008} assigned the 11.3 $\mu$m feature to SiC based on a slight bow in a low resolution (R$\sim$30) spectrum of SgrA IRS3. But the R$\sim$100 spectrum of the same target in \citet{Wright+2016} shows a distinct feature, which is much more consistent with crystalline silicates. The only detection of SiC based on the study of MIR polarization in the disk/envelope of a young star till date has been done by \citet{Fujiyoshi+2015} for SVS13 which has motivated us to investigate for evidence of silicon carbide (SiC) in the circumstellar environment of young stars.\\

Various studies have identified dust grains to be porous and fluffy in nature occurring as conglomerates comprising of multiple very tiny grains attached together due to collisions or interaction among the grains and various other processes which are likely to make the dust grains non-spherical and inhomogeneous [\citet{Gupta+2016}; and references therein]. In the absence of a precise theory for the study of composite dust grain properties, we take the help of various approximation techniques, e.g. the EMA: Effective Medium Approximation and the DDA: Discrete Dipole Approximation, to model the scattering by such grains. The EMA method [\citet{Bohren+1983}] uses T-Matrix which is based on Mie theory in order to determine the optical properties of small composite particles which may be either spherical or non-spherical in nature. However, in case the particle is inhomogeneous, EMA replaces it by a homogeneous particle with a single averaged optical property (refractive index, dielectric constant) which is also its limitation [\citet{Wolff+1994}; \citet{Ossenkopf+1991}; \citet{Perrin+1990}]. On the other hand, DDA [\citet{Purcell+1973}] is more computationally precise and works very well for inhomogeneous particles taking into consideration the effects of various irregularities (shape, surface roughness, internal structure of grains). DDA represents an arbitrarily shaped composite dust grain in an array form with dipole elements which will experience a polarization whenever an electromagnetic radiation is incident and also due to oscillation of the rest of the dipoles. The interaction and superposition of these two polarization components cause scattering cross sections and extinction [\citet{Draine+1988}]. In spite of the drawbacks of the EMA T-Matrix method, it is very convenient for large size parameters and large complex refractive index where DDA poses a computational challenge. In addition, EMA also allows the user to explore the suitability of various dust mixtures [\citet{Voshchinnikov+2006}; \citet{Saija+2001}]. Both these approximation methods have been used recently in a work by \citet{Gupta+2016} to model spheroidal composite dust grains made of silicate as the host material with silicon carbide (SiC) and graphite as inclusions. They have studied the effects of change in axial ratio, size, porosity and inclusions to the absorption efficiencies of dust grains in the 5-25 $\mu$m range with particular interest towards the 10 $\mu$m silicate feature observed in the MIR.\\

In this work, the presence of SiC in protoplanetary disks around young stars has been probed by studying the effects of grain shape, size, composition and volume fraction of inclusions on the polarization profile in the 8-13 $\mu$m range. This paper has been divided into the following sections: (2) Composite dust grain models (3) Comparison with observed data (4) Discussion and conclusions.

\section{Composite Dust Grain Models}

We have calculated the composite dust grain absorption efficiencies in the 8-13 $\mu$m range using DDA and EMA based models. We have used DDSCAT version 6.1 developed by \citet{Draine+Flatau+2003} (modified by \citet{Gupta+2005}) and the T-Matrix code developed by \citet{Mishchenko+2002} (modified by \citet{Vaidya+2009}) for our model calculations. The detailed analysis of various composite grain models is provided in \citet{Gupta+2016}. In this work, we have varied the following parameters to obtain the absorption efficiencies and subsequently the linear polarization:\\

\begin{figure}
\centering
	\includegraphics[width=8cm, height=7cm]{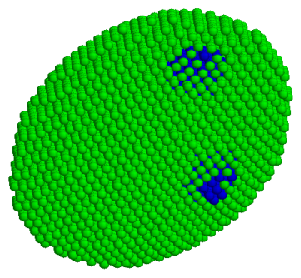}
    \includegraphics[width=5cm, height=5cm]{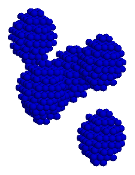}
    \caption{An oblate spheroidal composite dust grain model where the host material (silicate) is shown in green having an axial ratio (AR) = 1.33 with f = 0.1 (inclusions by volume). The inclusions are shown as blue implants in the upper panel and separately in magnified form in the lower panel. We are able to see only those inclusions in the upper panel which are located towards the perimeter of the spheroid while the rest are obstructed from our view.}
    \label{dust_model}
\end{figure}

\begin{description}

\item \textbf{Dust composition:} For our composite model calculations, the host spheroid has been set to be made up of silicates with inclusions of either silicon carbide (SiC) or graphite (Gr) at a time. The sites external to the grain are assumed as vacuum and internal sites are given to the silicate host. Although C-rich AGB stars are the primary source for SiC production [\citet{Speck+1997}], it has also been detected in interplanetary dust particles (IDPs) and meteorites [\citet{Bernatowicz+1987}; \citet{Bradley+2010}]. More recently, \citet{Fujiyoshi+2015} have identified SiC to be a major component of the protoplanetary disk/envelope of dust around binary star system SVS13. \citet{Fujiyoshi+2015} have modelled the polarization in SVS13 using amorphous silicates with SiC inclusions which motivated us to include SiC for our calculations as well. Graphite has been used since it is one of the prime candidates for the 2175 \AA \hspace{0.1cm} ultraviolet bump [\citet{Draine+1988}] observed in the extinction curve of the Galaxy. \citet{Alonso-Albi+2009} have successfully used a silicate/graphite mixture to model the dust in protoplanetary disks in a sample of Herbig Be stars. A significant amount of amorphous carbon or graphitic material is also found in the diffuse ISM ($\sim$30 \% by mass) in addition to the abundance of amorphous silicates. Since studies [\citet{Kohler+2004}; \citet{Voshchinnikov+2005}] have found dust grains to be fluffy and porous in nature, we have taken porous silicate based dust grains into consideration as well for our models. In addition, IDPs have also been found to consist of porous aggregates [\citet{Weidenschilling+2000}; \citet{Bradley+2003}; \citet{Poppe+2003}]. \\

\item \textbf{Axial ratio (AR):} We have used oblate (axial ratio $>$ 1) spheroids since interstellar extinction curves produced using models with grains in the form of oblate spheroids have been seen to show best fits with observed data [\citet{Gupta+2005}]. Good fits have been observed by \citet{Henning+1993} across 10 $\mu$m and by \citet{Kim+1995} in the near-IR using such models. In fact, dust grains are required to be non-spherical in order to produce linear polarization via dichroism [\citet{Draine+1984}; \citet{Aitken+1989}]. Recently, \citet{Fujiyoshi+2015} have used oblate spheroids with axial ratio 2 to successfully model the protoplanetary disk polarization in SVS13. Hence, we have used three values of the axial ratio (AR) for our spheroidal grains, AR = 1.33, 2.0 and 1.5 which correspond to  N = 9640, 14440 and 25896 dipoles respectively.\\

\item \textbf{Grain radius \& size integration:} \citet{Fujiyoshi+2015} had considered a grain radius of up to 0.5 $\mu$m for protoplanetary disks in the MIR based on the Rayleigh approximation (grain size $\ll$ wavelength). \citet{Min+2007} mention that the actual size of a particle is not important to determine the spectral shape in absorption/emission as long as it is in the Rayleigh domain. Hence, the absorption efficiencies have been calculated within the Rayleigh approximation using a grain size, a$_{min}$ = 0.1 $\mu$m to a$_{max}$ = 0.5 $\mu$m, where `a' represents the radius of a sphere such that its volume is equal to that of the dust grain spheroid. The details of particle size distribution are provided in \citet{Gupta+2016}.\\

\item \textbf{Volume fraction of inclusions:} \citet{Alonso-Albi+2009} have made use of a grain mixture comprising of 86\% silicate and 14\% graphite for protoplanetary disks around Herbig Be stars. In this work, we have used three different fractions by volume, f = 0.1 (10$\%$), 0.2 (20$\%$) and 0.3 (30$\%$) for SiC, graphite as well as porous grain inclusions (in compliance with the findings of \citet{Alonso-Albi+2009} and \citet{Gupta+2016}). A composite grain model with AR = 1.33, i.e. N = 9640 dipoles and having f = 0.1 (90\% silicates with 10\% inclusions) is illustrated in Figure \ref{dust_model}.\\

The amorphous/astronomical silicate and graphite refractive indices used in our dust models have been obtained from \citet{Draine+2003apj} while the SiC optical constants have been taken from \citet{Pegourie+1988}.

\end{description}

\section{Comparison with Observed Data}

\citet{LI+2018} have studied the magnetic fields of protoplanetary disks by measuring the polarization of pre-main sequence stars using CanariCam, a high-resolution MIR imaging instrument (7.5-25 $\mu$m) mounted over the 10.4 metres reflecting telescope, Gran Telescopio Canarias (GTC), situated at La Palma, Spain [\citet{Telesco+2003}; \citet{Packham+2005}]. Observations made by them at 8.7, 10.3 and 12.5 $\mu$m were selected such that they could sample the 10 $\mu$m silicate polarization feature which is ideally what we are trying to do from our composite grain models. They have separated the emission and absorption components of the star sample depending on the difference in polarization spectra across the 10 $\mu$m peak. The details of observation and data reduction using CanariCam have been presented in \citet{Zhang+2017b,Zhang+2017a} and \citet{LI+2018}. Two Herbig Be stars and one T-Tauri star showing polarization in absorption have been considered for this work from the sample of observations made by \citet{LI+2018} to compare with the linear polarization that we have calculated from the composite grain models.\\ 

MWC 1080A and MWC 297 are massive Herbig Be stars while HL Tau is a T-Tauri star which is the closest in our sample at 140 pc right before MWC 297 which is at a distance of 250 pc [\citet{LI+2018}; and references therein]. MWC 1080A is comparatively a very young star [\citet{Li+2014}] but all three show significant polarization in absorption. The two Herbig Be stars are surrounded by massive envelopes and seem to be still associated with the remains of their original molecular clouds [\citet{Alonso-Albi+2009}; \citet{Li+2014}; \citet{LI+2018}]. Even HL Tau is known to be enclosed by a torus shaped dusty envelope [\citet{Menshchikov+1999}] and the polarization observed in absorption for these three stars has been attributed to their torus/envelopes surrounding the protoplanetary disks with all of them showing peak absorption in intensity and maximum polarization around 10.3 $\mu$m. While the polarization profiles for MWC 297 and HL Tau show a sharp rise at 10.3 $\mu$m with subsequent fall to at least half this value on either side of the peak (i.e. at 8.7 and 12.5 $\mu$m), the polarization value is significantly high towards 12.5 $\mu$m in case of MWC 1080A, i.e. there is a shift in peak value towards 11 $\mu$m which is seen in the presence of larger grains/crystalline particles [\citet{Meeus+2001}]. Some details of the three stars as compiled by \citet{LI+2018} are shown in Table \ref{table:stars}. The distances are given in parsec (pc), mass in terms of solar mass (M$_\odot$) and the age is given in million years (Myr). The absorption feature around 10 $\mu$m has been denoted as `Abs' in the last column of Table \ref{table:stars}. \\
 
\begin{table}
\centering
\caption{Details of the three stars considered for comparison with our composite dust grain models as compiled by \citet{LI+2018}.}
\label{table:stars}
\begin{tabular}{lcccc}
\hline
Object & Distance & Mass & Age & 10 $\mu$m Silicate \\
 & (pc) & (M$_\odot$) & (Myr) & Feature Type \\
\hline
MWC 1080A  & 1000 & 20.6 & 0.22 & Abs \\
MWC 297  & 250 & 10 & 1 & Abs \\
HL Tau  & 140 & 0.7 & 1 & Abs \\
\hline
\end{tabular}
\end{table} 

\begin{figure}
    \begin{center}    
    \includegraphics[width=9cm, height=8cm]{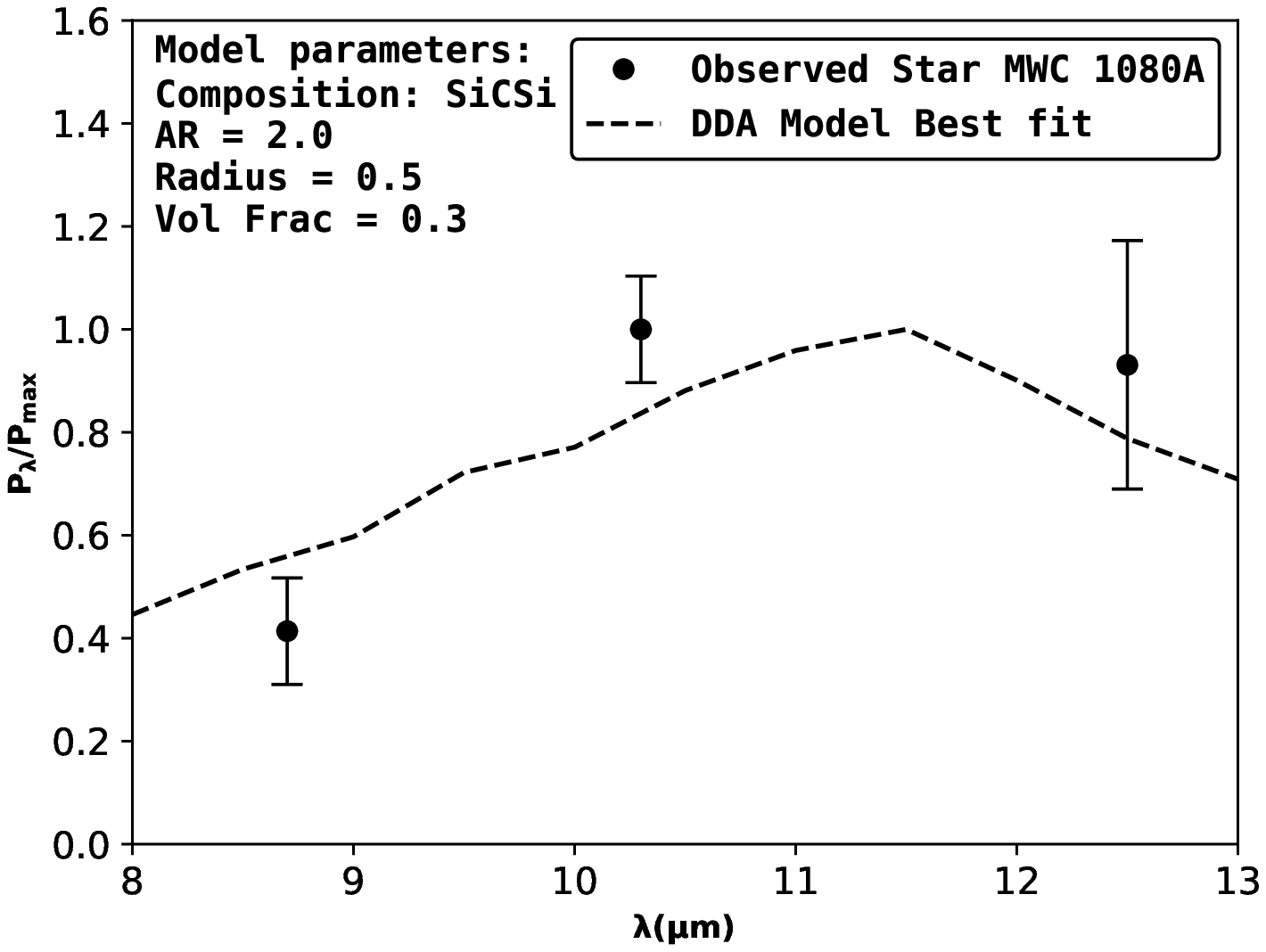}
    \includegraphics[width=9cm, height=8cm]{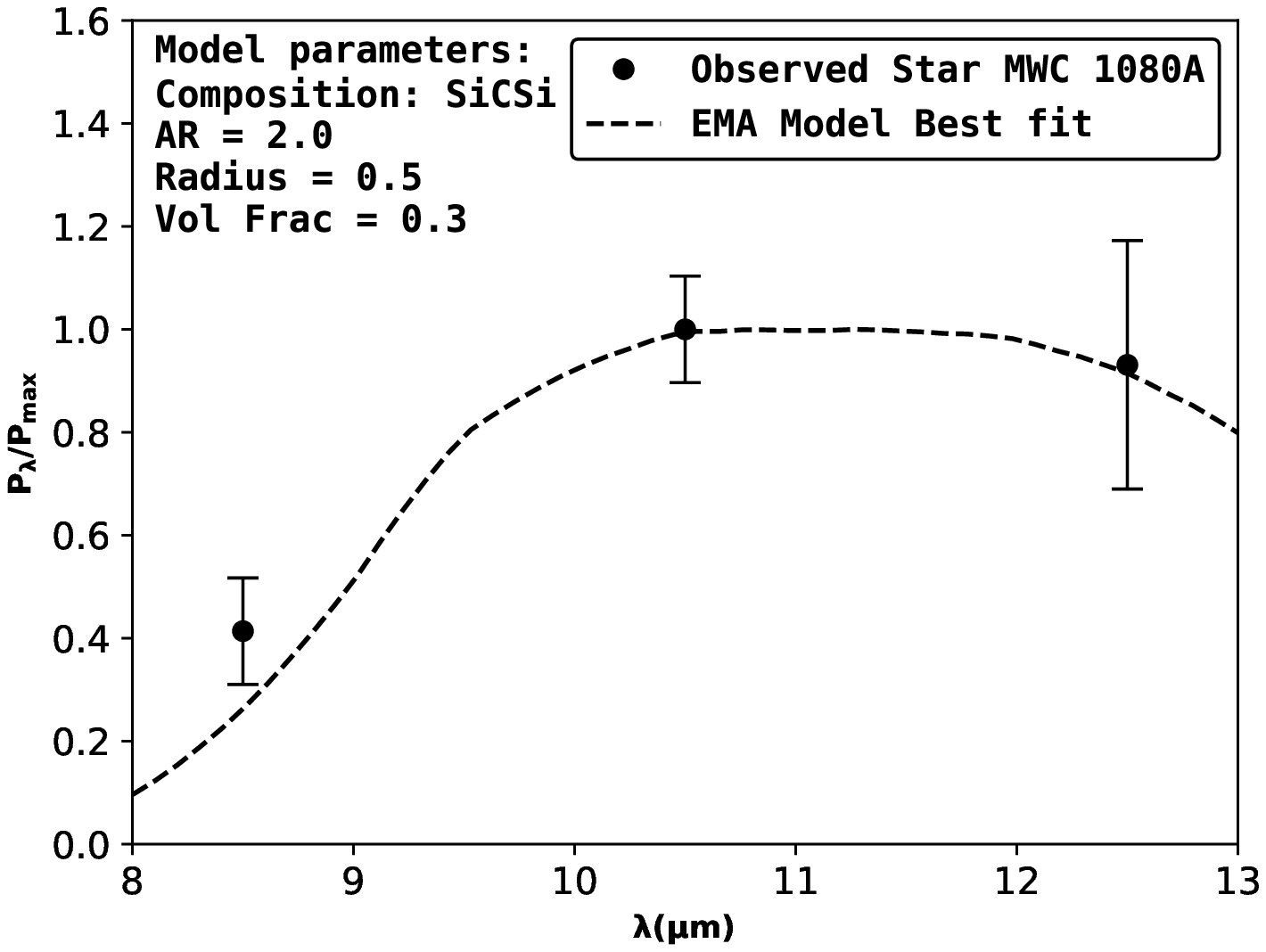}
    \caption{Composite grain models: DDA (top panel) and EMA (bottom panel) models with various combinations showing best fits to observed star MWC 1080A [\citet{LI+2018}].}
    \label{Fig:MW108}
    \end{center}
        \end{figure}

\begin{figure}
    \begin{center}    
    \includegraphics[width=9cm, height=8cm]{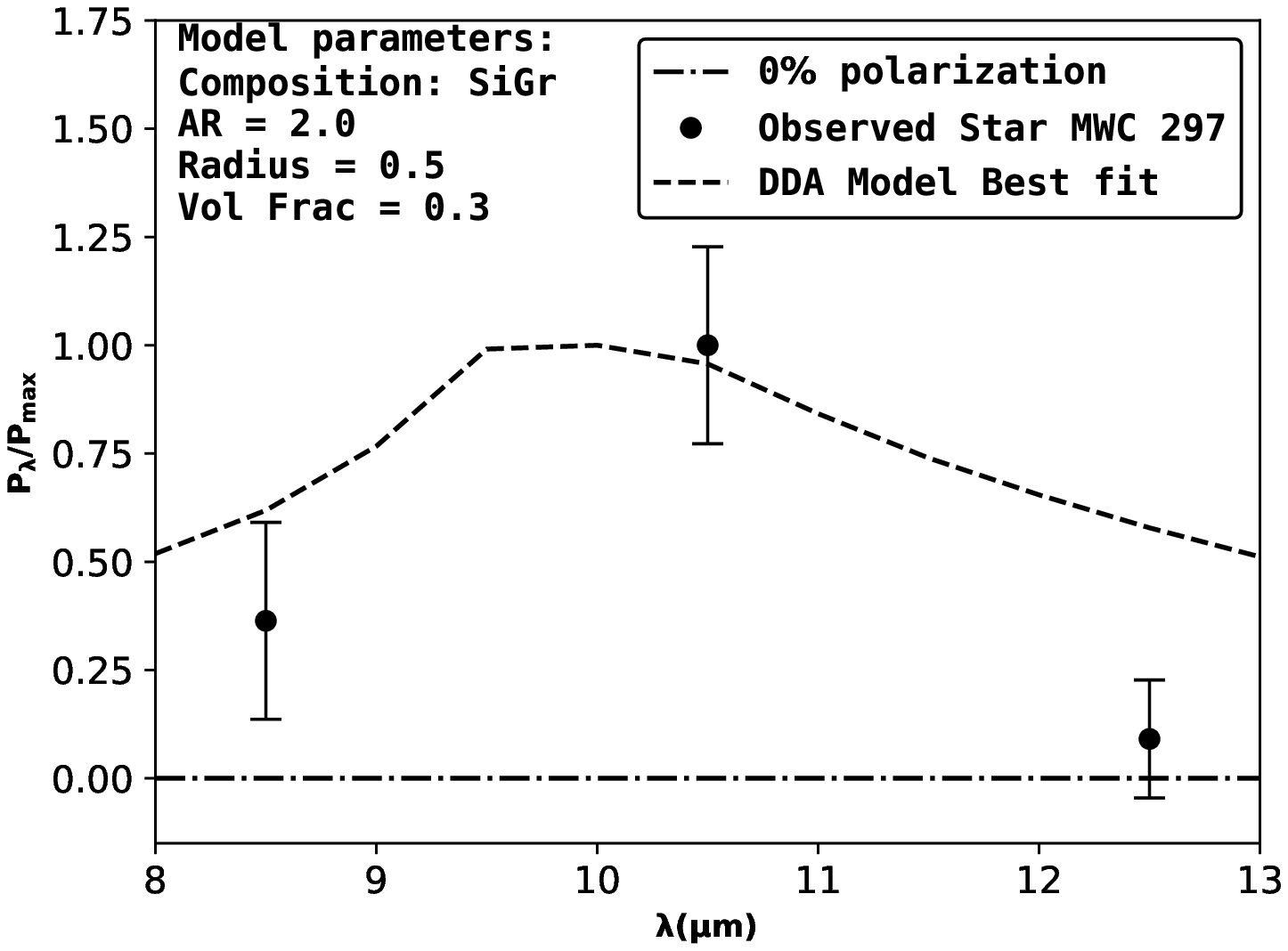}
    \includegraphics[width=9cm, height=8cm]{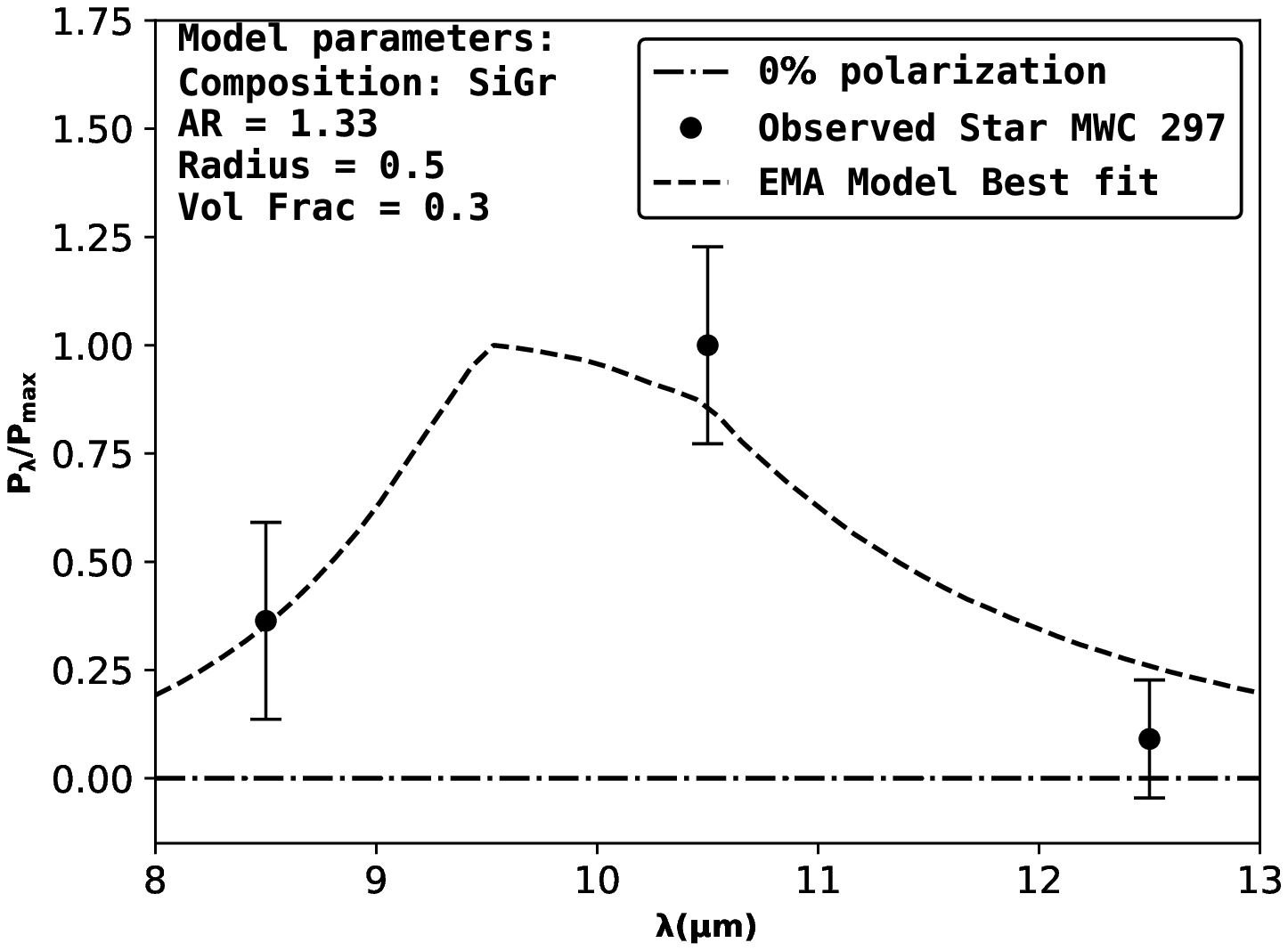}
    \caption{Composite grain models: DDA (top panel) and EMA (bottom panel) models with various combinations showing best fits to observed star MWC 297 [\citet{LI+2018}].}
    \label{Fig:MW297}
    \end{center}
        \end{figure}

\begin{figure}
    \begin{center}
    \includegraphics[width=9cm, height=8cm]{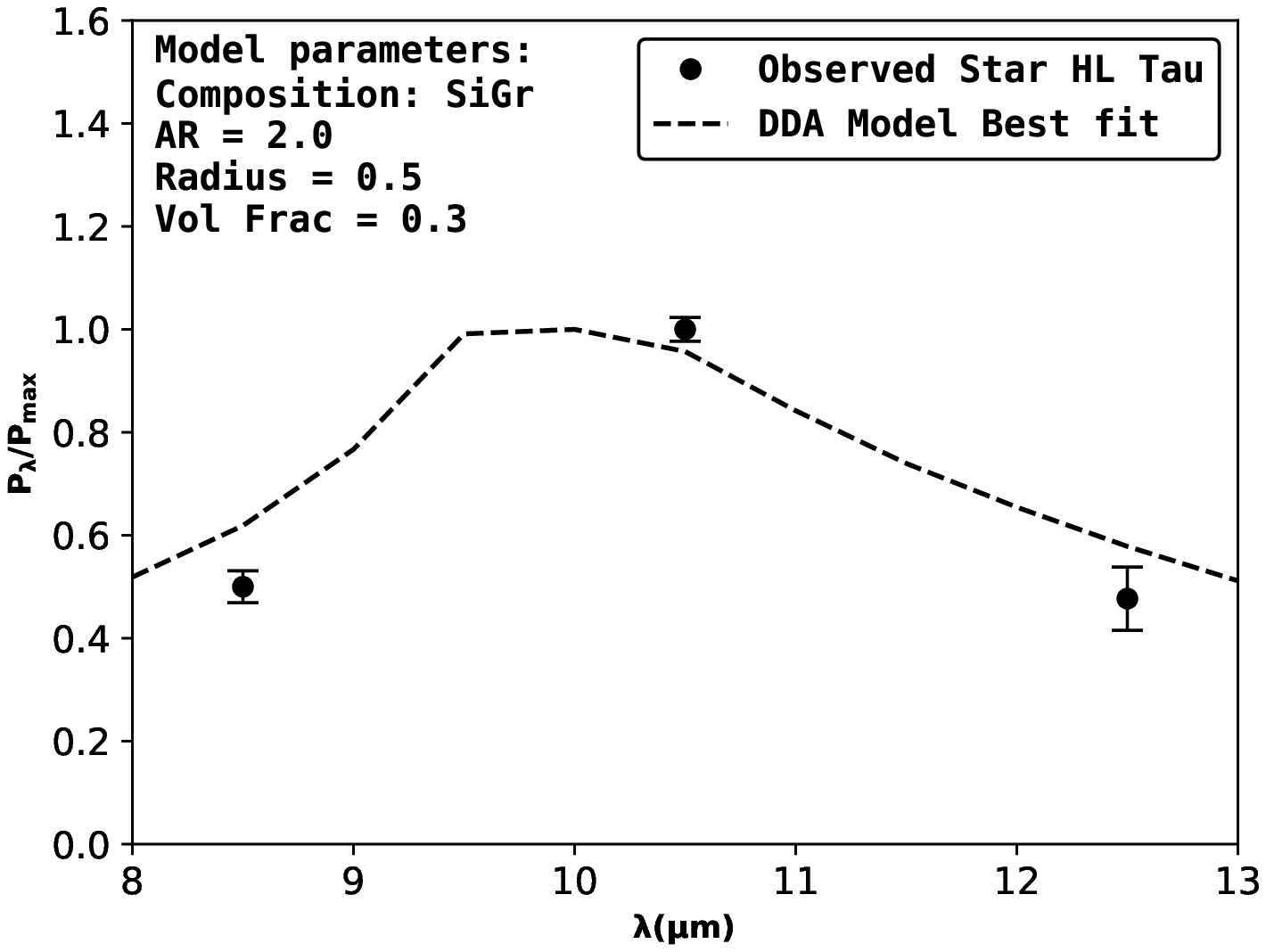}
    \includegraphics[width=9cm, height=8cm]{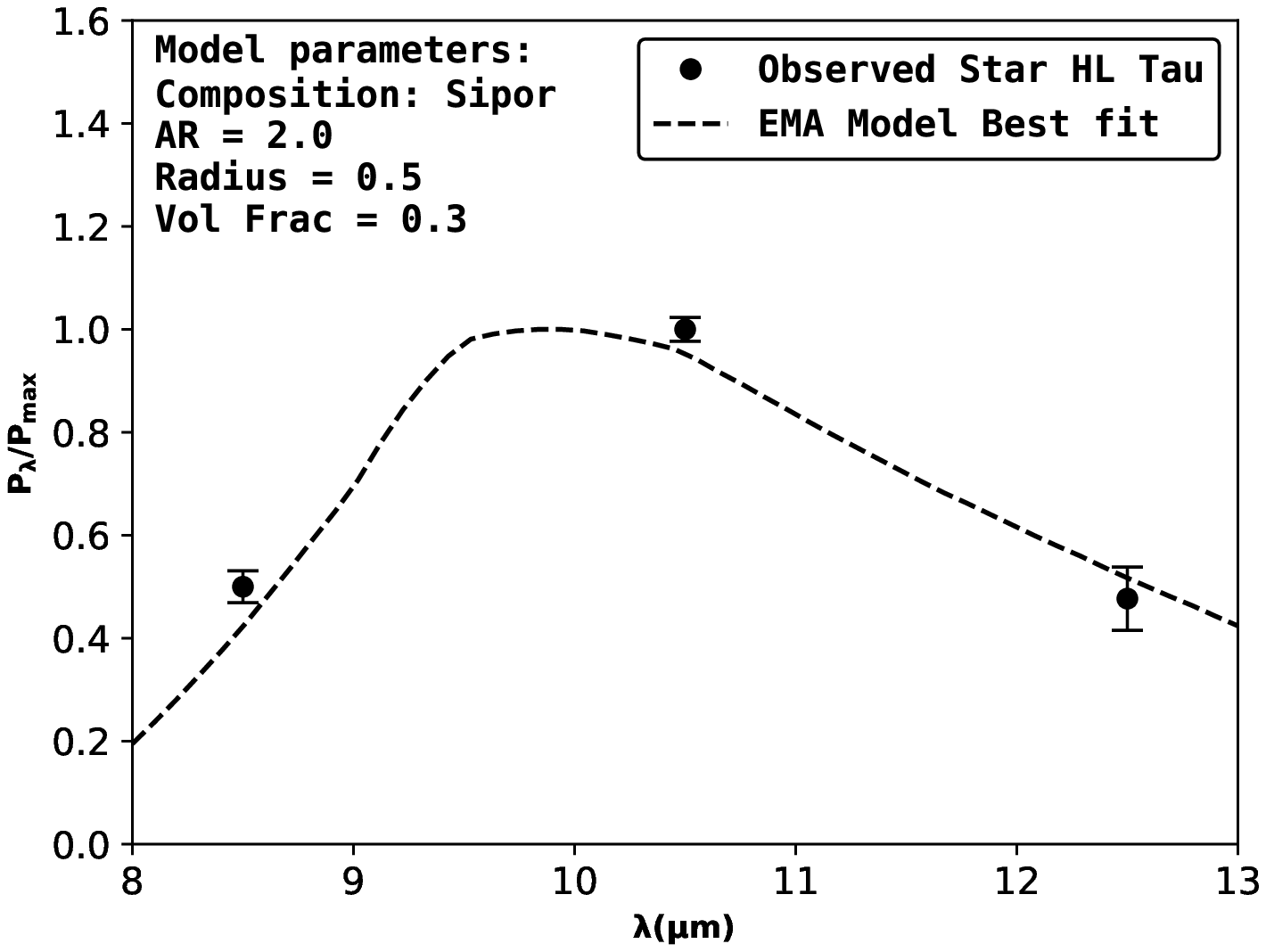}
    \caption{Composite grain models: DDA (top panel) and EMA (bottom panel) models with various combinations showing best fits to observed star HL Tau [\citet{LI+2018}].}
    \label{Fig:HLTau}
    \end{center}
        \end{figure}

In our composite dust grain models, it has been observed that the inclusion of SiC leads to a longward shift in peak while the inclusion of graphite leads to a shortward shift of peak wavelength from 10 $\mu$m (as detailed in \citet{Gupta+2016}). We have compared the polarization data for MWC 1080A, MWC 297 and HL Tau observed by \citet{LI+2018} with the calculated linear polarization from our models. The best fit values and details of various parameters considered in our models for each of the observed stars for DDA based calculations are presented in Table \ref{dda:parameters} and for EMA T-Matrix model calculations are presented in Table \ref{tmatrix:parameters}. We have shown the corresponding best fit plots for the three stars in Figures \ref{Fig:MW108}, \ref{Fig:MW297} and \ref{Fig:HLTau}.\\

\begin{table}
\centering
\caption{Details of best fit DDA based model parameters to the observed stars.}
\label{dda:parameters}
\begin{tabular}{ccccc}
\hline
Object & Material & AR & Size & Volume\\
& Composition &  & ($\mu$m) & Fraction\\
\hline
MWC 1080A & SiCSi & 2.0 & 0.5 & 0.3\\
MWC 297 & SiGr & 2.0 & 0.5 & 0.3  \\
HL Tau & SiGr & 2.0 & 0.5 & 0.3  \\
\hline
\end{tabular}
\end{table}

\begin{table}
\centering
\caption{Details of best fit EMA T-Matrix based model parameters to the observed stars.}
\label{tmatrix:parameters}
\begin{tabular}{ccccc}
\hline
Object & Material & AR & Size & Volume\\
& Composition &  & ($\mu$m) & Fraction \\
\hline
MWC 1080A & SiCSi & 2.0 & 0.5 & 0.3 \\
MWC 297 & SiGr & 1.33 & 0.5 & 0.3 \\
HL Tau & SiPor & 2.0 & 0.5 & 0.3 \\
\hline
\end{tabular}
\end{table}

\section{Discussion and Conclusions}

The results in Tables \ref{dda:parameters} and \ref{tmatrix:parameters} show that composite models comprising of silicate as the host with inclusions of SiC (f = 0.3) fit the observed MWC 1080A polarimetric data reasonably well for both DDA and EMA based models (Figure \ref{Fig:MW108}). On the other hand, the silicate/graphite mixture shows best fit to the observed MWC 297 polarization in both cases (Figure \ref{Fig:MW297}) while showing a good fit to HL Tau only in case of the DDA based model (Figure \ref{Fig:HLTau}). The porous silicate model shows a better fit to the HL Tau polarization when computed using the EMA method. The other exception we find in case of MWC 297 best fits is the difference in the axial ratio for DDA and EMA based models (AR = 2.0 and 1.33 respectively). The DDA approach is, of course, more reliable since it is rigorous and incorporates particle inhomogeneities which EMA fails to do as discussed earlier. \\

We find elongated spheroidal shaped grains (AR = 2.0) with an effective radius of 0.5 $\mu$m to best fit the observations in all three cases. As mentioned earlier, the grain size falls within the limits of Rayleigh approximation. Moreover, it has been observed that the dust grain size towards the inward regions of a protoplanetary disk is larger than those further out [\citet{Boekel+2008}]. \citet{Alonso-Albi+2009} state that the distance from a star could very well determine the dust grain growth rate as well as the grain composition. They have given models with a maximum grain radius, a$_{max}$ = 1 cm for grains in the midplane of the disks in MWC 1080A and MWC 297 and a$_{max}$ = 100 $\mu$m and 1 $\mu$m for grains in the disk surface of each star respectively. However, it is known that HL Tau [\citet{Beckwith+1990}] as well as both these Herbig Be stars are associated with massive envelopes [\citet{Alonso-Albi+2009}; \citet{Li+2014}; \citet{LI+2018}] which seem to be originators of the MIR absorptive polarization. If dust grain growth is indeed dependent on the distance from a star, the nature and size of dust grains found in such envelopes will resemble smaller ISM dust grains more closely than larger sized grains formed as a result of dust coagulation in the disks. In fact, it has been assumed that the dust grain growth from submicron to micron size takes place slowly over the course of a few million years [\citet{Dullemond+2008}]. Since MWC 297 and HL Tau are around 1 Myr in age (Table \ref{table:stars}), it can be safely assumed that the typical dust grains in their envelope are still sub-micron sized. This assumption fits even better to the case of MWC 1080A which is still at a very young age of 0.22 Myr (Table \ref{table:stars}).\\

In addition to the larger grain size, the flatness/shift of the typical 10 $\mu$m feature can also be caused due to the existence of crystalline silicate grains [\citet{Honda+2003}; \citet{Dullemond+2008}]. In such cases, although the grain size is small, a flat feature may be caused due to summing up of all the crystalline peaks, hence giving an impression of a feature caused by an amorphous larger sized grain. We have observed a silicate/graphite mixture in the dusty envelope around MWC 297 (Figure \ref{Fig:MW297}) which agrees with the observations of a similar mixture in the MWC 297 disk made by \citet{Alonso-Albi+2009}. However, \citet{LI+2018} favour a model for the MWC 297 spectrum which incorporates combined polarised emission and absorption of silicate grains and it indeed fits the polarization data better than our model where we assume polarization only in absorption. We have observed a similar silicate/graphite mixture around HL Tau from our DDA based calculations (Figure \ref{Fig:HLTau}). In contrast, the EMA based calculations show a porous silicate model to better fit with HL Tau observations which is in line with the pure silicate absorption model put forward by \citet{LI+2018} for the same target. \\

The SiC and silicate mixture which we have observed for MWC 1080A does not agree with the surface/midplane dust models given by \citet{Alonso-Albi+2009} for a silicate/graphite mixture. The deviation in results could be attributed to their simple model with different grain size distribution and also to the MWC 1080A polarization observed by \citet{LI+2018} which seems to arise from the envelope rather than the disk surface/midplane. \citet{LI+2018} favour a pure absorption model with silicates only for MWC 1080A but it does not fit the 12.5 $\mu$m polarization data which our SiCSi model does. A change in polarization position angle (PA) is seen across the 8--13 $\mu$m spectrum for both MWC 1080A and MWC 297. For MWC 297, the change in PA is within the respective 8.7, 10.3 and 12.5 $\mu$m uncertainty bounds, while this is not the case for MWC 1080A. In most cases in the thermal MIR, such a change is related to the presence of combined polarised emission and absorption, as has been accounted for MWC 297 by \citet{LI+2018}, though it is feasible that a changing grain alignment and dust composition along the line-of-sight could also produce a variation. However, as noted by \citet{LI+2018}, MWC 1080A seems to be a special case and the polarization angle observed deviates from models, the value at 12.5 $\mu$m in particular, where data is also less statistically significant (about 4$\sigma$) than at 8.7 $\mu$m and 10.3 $\mu$m (10$\sigma$). While one reason for the inability of our model to account for the change in PA towards 12.5 $\mu$m could be that the assumption of polarization is only in absorption, there is also a possibility of bimodal grain size distribution [\citet{Dullemond+2008}] in MWC 1080A being a contributory factor towards such an observation of offset in PA, which our model does not take into consideration. \\

Such an occurrence of polarization due to absorption beyond the typical 10 $\mu$m feature was observed by \citet{Aitken+1988} which was considered to be due to the presence of crystalline silicates peaking at 11.2 $\mu$m. \citet{Min+2007} state that SiC grains that are irregular in shape show a broad feature in the spectra which peaks near 11.25 $\mu$m. Since we have indeed observed spheroidal silicate grains with AR = 2.0 having SiC inclusions to show such a broad feature peaking near 11.3 $\mu$m with significant polarization observed at 12.5 $\mu$m (Figure \ref{Fig:MW108}), we can assume our observed polarization in MWC 1080A to have been caused due to the SiC component present in the dust grains. The possible detection of SiC in the envelope of MWC 1080A here is very interesting due to the fact that there has only been one evidence of SiC detection while studying the MIR polarization in absorption which was made by \citet{Fujiyoshi+2015} for SVS13. However, SiC at the level of tens of percent volume fraction has a profound effect not just on the polarisation but also on the conventional spectrum as seen in the case of SVS13. Unfortunately, the conventional spectrum of MWC 1080A given by \citet{Sakon+2007} does not show any such apparent absorption feature around 11.2 $\mu$m which contradicts the findings of our model. \\

While most cases of crystalline silicate detection have been reported in emission [\citet{Watson+2009}; \citet{Olofson+2009}; \citet{Juhasz+2010}], recent studies have shown a considerable number of such features in absorption as well [\citet{Poteet+2011}; \citet{Fujiyoshi+2015}; \citet{Wright+2016}]. All three objects in our sample of study: MWC 1080A, MWC 297 and HL Tau, show polarization in absorption but only the MWC 1080A polarization seems to have been caused due to SiC mixed with silicates. The dust grains accountable for the crystalline silicate features in absorption seem to originate in outer regions which are cold [\citet{Fujiyoshi+2015}; \citet{Wright+2016}]. Since MWC 1080A is a very young but massive star with a small disk size, most of the observed polarization is caused due to the envelope which consists of material flowing in from the ISM. Moreover, multiple evidences have been found of the occurrence of crystalline silicates in cold outer shells/disks of Herbig stars [\citet{Suh+2011}] with separate mechanisms proposed for either transport/mixture of such dust to the outlying regions or for the formation of silicates in crystalline form in low temperature disks/envelopes due to shocks [\citet{Juhasz+2010}; \citet{Suh+2011}].\\

\begin{figure}
    \begin{center}
    \includegraphics[width=9cm, height=8cm]{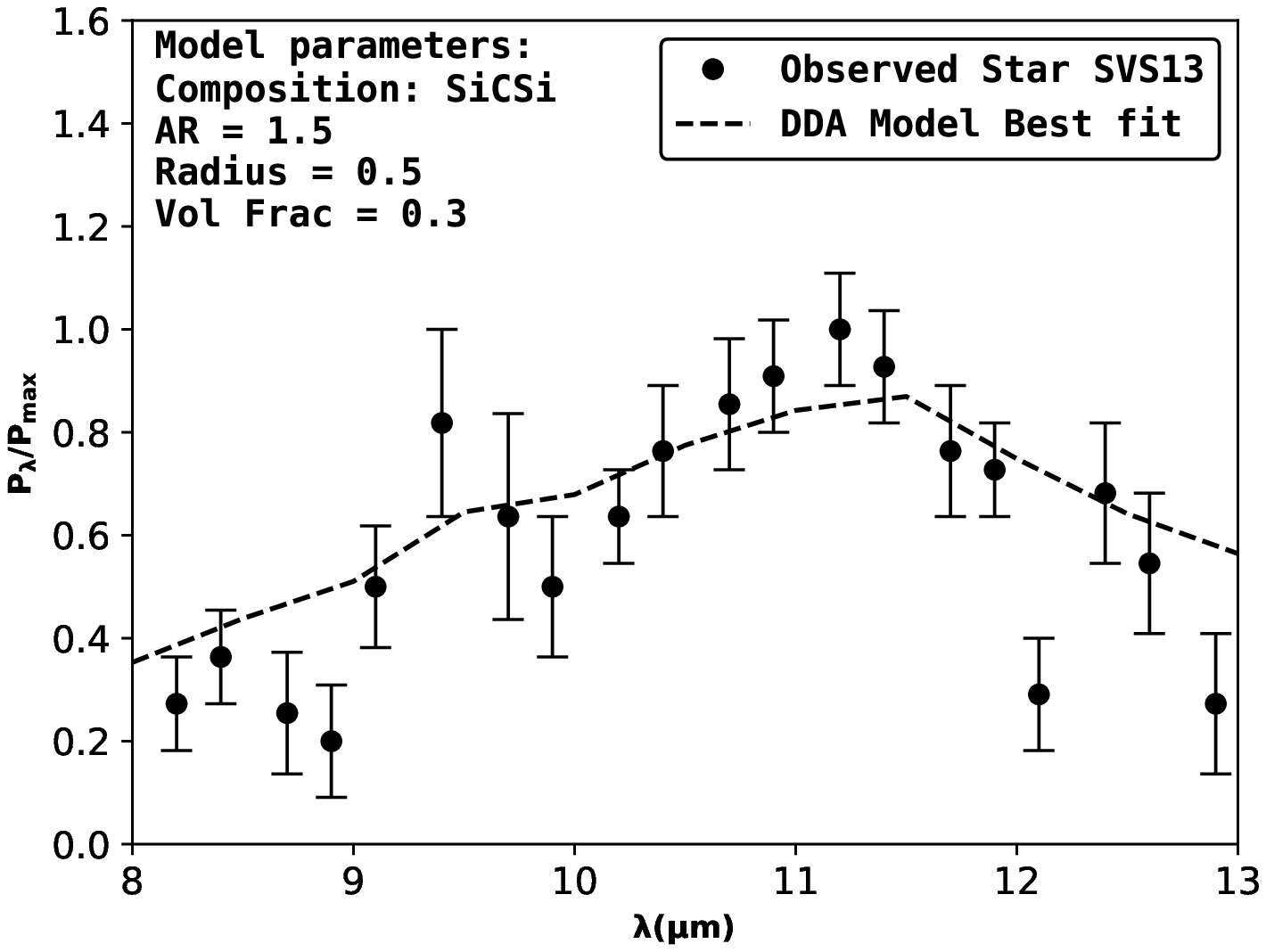}
    \includegraphics[width=9cm, height=8cm]{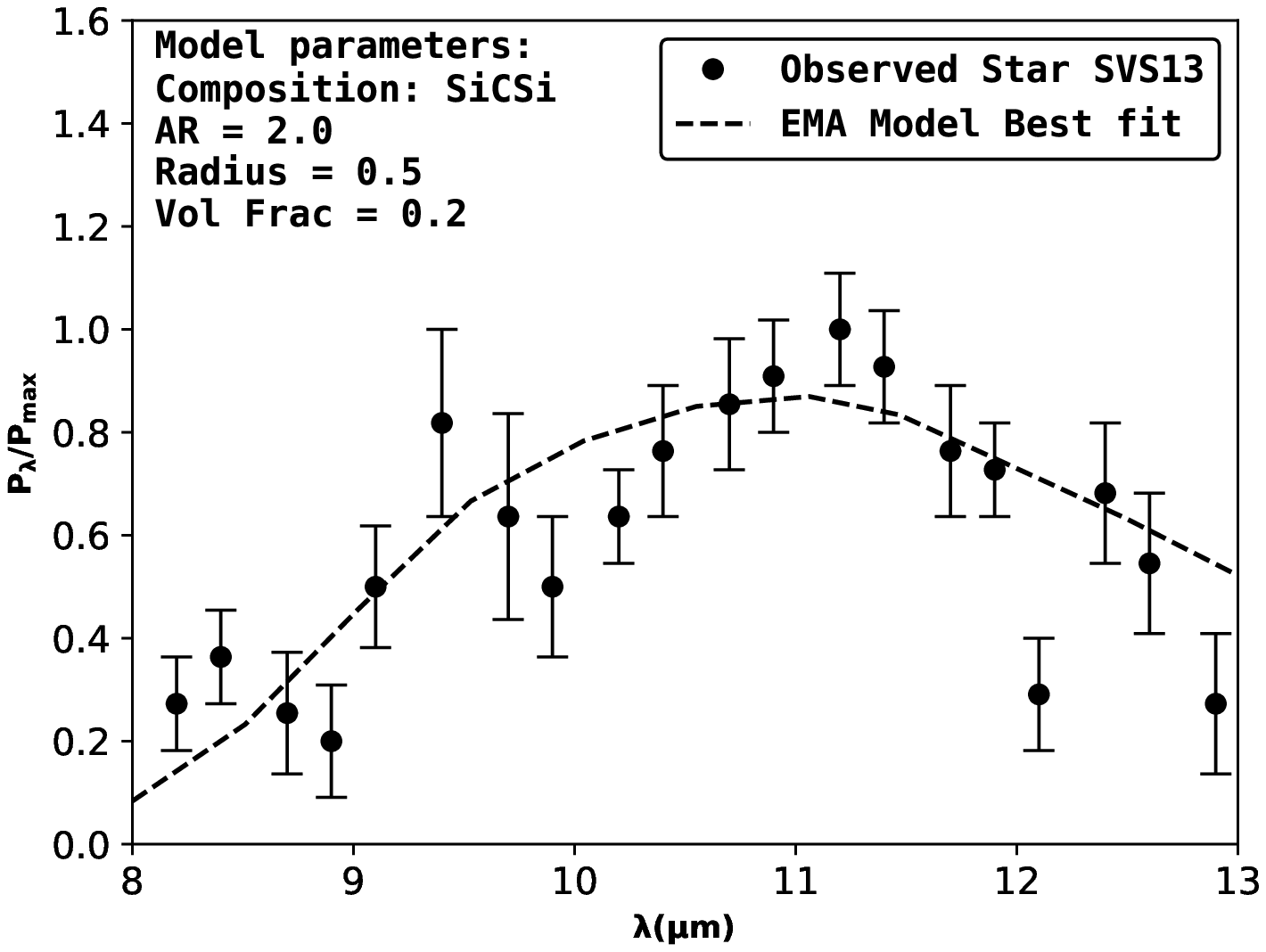}
    \caption{Composite grain models: DDA (top panel) and EMA (bottom panel) models with various combinations showing best fits to observed star SVS13 [\citet{Fujiyoshi+2015}].}
    \label{Fig:SVS13}
    \end{center}
        \end{figure}

We have also checked the validity of our composite dust models with the polarization data of SVS13 presented in \citet{Fujiyoshi+2015} as shown in Figure \ref{Fig:SVS13}. Our DDA based calculations give a model with SiCSi, having 0.3 volume fraction of inclusions, for oblate spheroidal dust grains (AR = 1.5) of effective radius 0.5 $\mu$m to best fit the SVS13 polarization data. Our EMA T-Matrix based calculations are in agreement to the type of inclusion, i.e. SiC, but vary in the degree of oblateness (AR = 2.0) and volume fraction of inclusions (0.2). \citet{Fujiyoshi+2015} had found similar results with an EMA based model having SiC inclusions which are in agreement with our models. They had found an AR = 2.0 and volume fraction of inclusions 0.25 as best fit parameters when using the refractive indices of $\alpha$-SiC given by \citet{Pegourie+1988}, which is the same one we have used here. The notable difference is only in the AR between our DDA based model and the EMA based results obtained by \citet{Fujiyoshi+2015} as supported by our calculations.\\

In conclusion, our composite dust grain models account very well for the occurrence of SiC in SVS13, in agreement with the findings of \citet{Fujiyoshi+2015}. The observed polarization in HL Tau and MWC 297 can also be accounted for within the constraints of our model parameters. The only exception here has been the case of MWC 1080A for which the SiCSi model (i) can account for the observed polarization spectrum, but (ii) does not simultaneously account for the very different polarization position angle at 12.5 $\mu$m as compared to 8.7 and 10.3 $\mu$m, and (iii) does not agree with a lack of absorption at 11.2 $\mu$m in the 8-13 $\mu$m conventional spectrum of MWC 1080A [\citet{Sakon+2007}]. \\

Our results have been limited by the lack of observed polarimetric data between 10.3 and 12.5 $\mu$m, the availability of which would add more credibility to the findings of this work. There is also a prospect of exploring the possible contribution of polycyclic aromatic hydrocarbon (PAH) molecules, which has already been detected for MWC 1080A [\citet{Zhang+2017a}], with the availability of reliable MIR polarization data for these young pre-main sequence objects.\\

\section*{Acknowledgements}

The authors are indebted to the reviewer, Christopher M. Wright, for his expert comments and suggestions which have greatly improved the content and presentation of this work. R. Gogoi is grateful to the IUCAA associateship programme for their support and hospitality.




\bibliographystyle{mnras}
\bibliography{mybibfile} 

\begin{thebibliography}{}
\makeatletter
\relax
\def\mn@urlcharsother{\let\do\@makeother \do\$\do\&\do\#\do\^\do\_\do\%\do\~}
\def\mn@doi{\begingroup\mn@urlcharsother \@ifnextchar [ {\mn@doi@}
  {\mn@doi@[]}}
\def\mn@doi@[#1]#2{\def\@tempa{#1}\ifx\@tempa\@empty \href
  {http://dx.doi.org/#2} {doi:#2}\else \href {http://dx.doi.org/#2} {#1}\fi
  \endgroup}
\def\mn@eprint#1#2{\mn@eprint@#1:#2::\@nil}
\def\mn@eprint@arXiv#1{\href {http://arxiv.org/abs/#1} {{\tt arXiv:#1}}}
\def\mn@eprint@dblp#1{\href {http://dblp.uni-trier.de/rec/bibtex/#1.xml}
  {dblp:#1}}
\def\mn@eprint@#1:#2:#3:#4\@nil{\def\@tempa {#1}\def\@tempb {#2}\def\@tempc
  {#3}\ifx \@tempc \@empty \let \@tempc \@tempb \let \@tempb \@tempa \fi \ifx
  \@tempb \@empty \def\@tempb {arXiv}\fi \@ifundefined
  {mn@eprint@\@tempb}{\@tempb:\@tempc}{\expandafter \expandafter \csname
  mn@eprint@\@tempb\endcsname \expandafter{\@tempc}}}

\bibitem[\protect\citeauthoryear{{Acke}, {Bouwman}, {van Winckel}, {Waters},
  {van Boekel}  \& {van den Ancker}}{{Acke} et~al.}{2005}]{Acke+2005}
{Acke} B.,  {Bouwman} J.,  {van Winckel} H.,  {Waters} R.,  {van Boekel} R.,
  {van den Ancker} M.,  2005, {Probing the disk mineralogy and geometry of
  Herbig Ae/Be stars}, Spitzer Proposal

\bibitem[\protect\citeauthoryear{{Aitken}}{{Aitken}}{1989}]{Aitken+1989}
{Aitken} D.~K.,  1989, in {B{\"o}hm-Vitense} E.,  ed.,  ESA Special Publication
  Vol. 290, Infrared Spectroscopy in Astronomy.

\bibitem[\protect\citeauthoryear{{Aitken}, {Roche}, {Smith}, {James}  \&
  {Hough}}{{Aitken} et~al.}{1988}]{Aitken+1988}
{Aitken} D.~K.,  {Roche} P.~F.,  {Smith} C.~H.,  {James} S.~D.,   {Hough}
  J.~H.,  1988, \mn@doi [Monthly Notices of the Royal Astronomical Society]
  {10.1093/mnras/230.4.629}, \href
  {http://adsabs.harvard.edu/abs/1988MNRAS.230..629A} {230, 629}

\bibitem[\protect\citeauthoryear{{Alonso-Albi}, {Fuente}, {Bachiller}, {Neri},
  {Planesas}, {Testi}, {Bern{\'e}}  \& {Joblin}}{{Alonso-Albi}
  et~al.}{2009}]{Alonso-Albi+2009}
{Alonso-Albi} T.,  {Fuente} A.,  {Bachiller} R.,  {Neri} R.,  {Planesas} P.,
  {Testi} L.,  {Bern{\'e}} O.,   {Joblin} C.,  2009, \mn@doi [Astronomy and
  Astrophysics] {10.1051/0004-6361/200810401}, \href
  {http://adsabs.harvard.edu/abs/2009A%26A...497..117A} {497, 117}

\bibitem[\protect\citeauthoryear{{Beckwith}, {Sargent}, {Chini}  \&
  {Guesten}}{{Beckwith} et~al.}{1990}]{Beckwith+1990}
{Beckwith} S.~V.~W.,  {Sargent} A.~I.,  {Chini} R.~S.,   {Guesten} R.,  1990,
  \mn@doi [Astronomical Journal] {10.1086/115385}, \href
  {http://adsabs.harvard.edu/abs/1990AJ.....99..924B} {99, 924}

\bibitem[\protect\citeauthoryear{{Bernatowicz}, {Fraundorf}, {Ming}, {Anders},
  {Wopenka}, {Zinner}  \& {Fraundorf}}{{Bernatowicz}
  et~al.}{1987}]{Bernatowicz+1987}
{Bernatowicz} T.,  {Fraundorf} G.,  {Ming} T.,  {Anders} E.,  {Wopenka} B.,
  {Zinner} E.,   {Fraundorf} P.,  1987, \mn@doi [Nature] {10.1038/330728a0},
  \href {http://adsabs.harvard.edu/abs/1987Natur.330..728B} {330, 728}

\bibitem[\protect\citeauthoryear{{Bernatowicz}, {Croat}  \&
  {Daulton}}{{Bernatowicz} et~al.}{2006}]{Bernatowicz+2006}
{Bernatowicz} T.~J.,  {Croat} T.~K.,   {Daulton} T.~L.,  2006, {Origin and
  Evolution of Carbonaceous Presolar Grains in Stellar Environments}.
pp 109--126

\bibitem[\protect\citeauthoryear{{Bohren} \& {Huffman}}{{Bohren} \&
  {Huffman}}{1983}]{Bohren+1983}
{Bohren} C.~F.,  {Huffman} D.~R.,  1983, {Absorption and scattering of light by
  small particles}

\bibitem[\protect\citeauthoryear{{Bouwman} et~al.,}{{Bouwman}
  et~al.}{2008}]{Bouman+2008}
{Bouwman} J.,  et~al., 2008, \mn@doi [The Astrophysical Journal]
  {10.1086/587793}, \href {http://adsabs.harvard.edu/abs/2008ApJ...683..479B}
  {683, 479}

\bibitem[\protect\citeauthoryear{{Bradley}}{{Bradley}}{2003}]{Bradley+2003}
{Bradley} J.~P.,  2003, \mn@doi [Treatise on Geochemistry]
  {10.1016/B0-08-043751-6/01152-X}, \href
  {http://adsabs.harvard.edu/abs/2003TrGeo...1..689B} {1, 711}

\bibitem[\protect\citeauthoryear{{Bradley}}{{Bradley}}{2010}]{Bradley+2010}
{Bradley} J.,  2010, in {Henning} T.,  ed.,  Lecture Notes in Physics, Berlin
  Springer Verlag Vol. 815, Lecture Notes in Physics, Berlin Springer Verlag.
  pp 259--276, \mn@doi{10.1007/978-3-642-13259-9_6}

\bibitem[\protect\citeauthoryear{{Draine}}{{Draine}}{1988}]{Draine+1988}
{Draine} B.~T.,  1988, \mn@doi [The Astrophysical Journal] {10.1086/166795},
  \href {http://adsabs.harvard.edu/abs/1988ApJ...333..848D} {333, 848}

\bibitem[\protect\citeauthoryear{Draine}{Draine}{2003}]{Draine+2003apj}
Draine B.,  2003, The Astrophysical Journal, 598, 1017

\bibitem[\protect\citeauthoryear{{Draine} \& {Flatau}}{{Draine} \&
  {Flatau}}{2003}]{Draine+Flatau+2003}
{Draine} B.~T.,  {Flatau} P.~J.,  2003, ArXiv Astrophysics e-prints, \href
  {http://adsabs.harvard.edu/abs/2003astro.ph..9069D} {}

\bibitem[\protect\citeauthoryear{{Draine} \& {Lee}}{{Draine} \&
  {Lee}}{1984}]{Draine+1984}
{Draine} B.~T.,  {Lee} H.~M.,  1984, \mn@doi [The Astrophysical Journal]
  {10.1086/162480}, \href {http://adsabs.harvard.edu/abs/1984ApJ...285...89D}
  {285, 89}

\bibitem[\protect\citeauthoryear{{Dullemond} \& {Dominik}}{{Dullemond} \&
  {Dominik}}{2008}]{Dullemond+2008}
{Dullemond} C.~P.,  {Dominik} C.,  2008, \mn@doi [Astronomy and Astrophysics]
  {10.1051/0004-6361:200809745}, \href
  {http://adsabs.harvard.edu/abs/2008A%26A...487..205D} {487, 205}

\bibitem[\protect\citeauthoryear{Floss, Stadermann, Kearsley, Burchell  \&
  Ong}{Floss et~al.}{2013}]{Floss+2013}
Floss C.,  Stadermann F.~J.,  Kearsley A.~T.,  Burchell M.~J.,   Ong W.,  2013,
  The Astrophysical Journal, 763, 140

\bibitem[\protect\citeauthoryear{{Fujiyoshi}, {Wright}  \& {Moore}}{{Fujiyoshi}
  et~al.}{2015}]{Fujiyoshi+2015}
{Fujiyoshi} T.,  {Wright} C.~M.,   {Moore} T.~J.~T.,  2015, \mn@doi [Monthly
  Notices of the Royal Astronomical Society] {10.1093/mnras/stv1171}, \href
  {http://adsabs.harvard.edu/abs/2015MNRAS.451.3371F} {451, 3371}

\bibitem[\protect\citeauthoryear{{Gupta}, {Mukai}, {Vaidya}, {Sen}  \&
  {Okada}}{{Gupta} et~al.}{2005}]{Gupta+2005}
{Gupta} R.,  {Mukai} T.,  {Vaidya} D.~B.,  {Sen} A.~K.,   {Okada} Y.,  2005,
  \mn@doi [Astronomy and Astrophysics] {10.1051/0004-6361:20052707}, \href
  {http://adsabs.harvard.edu/abs/2005A%26A...441..555G} {441, 555}

\bibitem[\protect\citeauthoryear{{Gupta}, {Vaidya}  \& {Dutta}}{{Gupta}
  et~al.}{2016}]{Gupta+2016}
{Gupta} R.,  {Vaidya} D.~B.,   {Dutta} R.,  2016, \mn@doi [Monthly Notices of
  the Royal Astronomical Society] {10.1093/mnras/stw1710}, \href
  {http://adsabs.harvard.edu/abs/2016MNRAS.462..867G} {462, 867}

\bibitem[\protect\citeauthoryear{{Henning}}{{Henning}}{2010}]{Henning+2010}
{Henning} T.,  2010, \mn@doi [Annual Review of Astronomy and Astrophysics]
  {10.1146/annurev-astro-081309-130815}, \href
  {http://adsabs.harvard.edu/abs/2010ARA%26A..48...21H} {48, 21}

\bibitem[\protect\citeauthoryear{{Henning} \& {Stognienko}}{{Henning} \&
  {Stognienko}}{1993}]{Henning+1993}
{Henning} T.,  {Stognienko} R.,  1993, Astronomy and Astrophysics, \href
  {http://adsabs.harvard.edu/abs/1993A%26A...280..609H} {280, 609}

\bibitem[\protect\citeauthoryear{{Herbig}}{{Herbig}}{1960}]{Herbig+1960}
{Herbig} G.~H.,  1960, \mn@doi [The Astrophysical Journal] {10.1086/146876},
  \href {http://adsabs.harvard.edu/abs/1960ApJ...131..632H} {131, 632}

\bibitem[\protect\citeauthoryear{{Hofmeister}, {Pitman}, {Goncharov}  \&
  {Speck}}{{Hofmeister} et~al.}{2009}]{Hofmeister+2009}
{Hofmeister} A.~M.,  {Pitman} K.~M.,  {Goncharov} A.~F.,   {Speck} A.~K.,
  2009, \mn@doi [The Astrophysical Journal] {10.1088/0004-637X/696/2/1502},
  \href {http://adsabs.harvard.edu/abs/2009ApJ...696.1502H} {696, 1502}

\bibitem[\protect\citeauthoryear{{Honda}, {Kataza}, {Okamoto}, {Miyata},
  {Yamashita}, {Sako}, {Takubo}  \& {Onaka}}{{Honda} et~al.}{2003}]{Honda+2003}
{Honda} M.,  {Kataza} H.,  {Okamoto} Y.~K.,  {Miyata} T.,  {Yamashita} T.,
  {Sako} S.,  {Takubo} S.,   {Onaka} T.,  2003, \mn@doi [The Astrophysical
  Journal Letters] {10.1086/374034}, \href
  {http://adsabs.harvard.edu/abs/2003ApJ...585L..59H} {585, L59}

\bibitem[\protect\citeauthoryear{{Juh{\'a}sz} et~al.,}{{Juh{\'a}sz}
  et~al.}{2010}]{Juhasz+2010}
{Juh{\'a}sz} A.,  et~al., 2010, \mn@doi [The Astrophysical Journal]
  {10.1088/0004-637X/721/1/431}, \href
  {http://adsabs.harvard.edu/abs/2010ApJ...721..431J} {721, 431}

\bibitem[\protect\citeauthoryear{{Kemper}, {Vriend}  \& {Tielens}}{{Kemper}
  et~al.}{2004}]{Kemper+2004}
{Kemper} F.,  {Vriend} W.~J.,   {Tielens} A.~G.~G.~M.,  2004, \mn@doi [The
  Astrophysical Journal] {10.1086/421339}, \href
  {http://adsabs.harvard.edu/abs/2004ApJ...609..826K} {609, 826}

\bibitem[\protect\citeauthoryear{{Kemper}, {Vriend}  \& {Tielens}}{{Kemper}
  et~al.}{2005}]{Kemper+2005}
{Kemper} F.,  {Vriend} W.~J.,   {Tielens} A.~G.~G.~M.,  2005, \mn@doi [The
  Astrophysical Journal] {10.1086/447764}, \href
  {http://adsabs.harvard.edu/abs/2005ApJ...633..534K} {633, 534}

\bibitem[\protect\citeauthoryear{{Kessler-Silacci} et~al.,}{{Kessler-Silacci}
  et~al.}{2006}]{Kessler-Silacci+2006}
{Kessler-Silacci} J.,  et~al., 2006, \mn@doi [The Astrophysical Journal]
  {10.1086/499330}, \href {http://adsabs.harvard.edu/abs/2006ApJ...639..275K}
  {639, 275}

\bibitem[\protect\citeauthoryear{{Kim} \& {Martin}}{{Kim} \&
  {Martin}}{1995}]{Kim+1995}
{Kim} S.-H.,  {Martin} P.~G.,  1995, \mn@doi [The Astrophysical Journal]
  {10.1086/175604}, \href {http://adsabs.harvard.edu/abs/1995ApJ...444..293K}
  {444, 293}

\bibitem[\protect\citeauthoryear{{K{\"o}hler} \& {Mann}}{{K{\"o}hler} \&
  {Mann}}{2004}]{Kohler+2004}
{K{\"o}hler} M.,  {Mann} I.,  2004, \mn@doi [Journal of Quantitative
  Spectroscopy and Radiative Transfer] {10.1016/j.jqsrt.2004.05.003}, \href
  {http://adsabs.harvard.edu/abs/2004JQSRT..89..453K} {89, 453}

\bibitem[\protect\citeauthoryear{{Leinert}, {Haas}, {{\'A}brah{\'a}m}  \&
  {Richichi}}{{Leinert} et~al.}{2001}]{Leinert+2001}
{Leinert} C.,  {Haas} M.,  {{\'A}brah{\'a}m} P.,   {Richichi} A.,  2001,
  \mn@doi [Astronomy and Astrophysics] {10.1051/0004-6361:20010904}, \href
  {http://adsabs.harvard.edu/abs/2001A%26A...375..927L} {375, 927}

\bibitem[\protect\citeauthoryear{{Leinert} et~al.,}{{Leinert}
  et~al.}{2004}]{Leinert+2004}
{Leinert} C.,  et~al., 2004, \mn@doi [Astronomy and Astrophysics]
  {10.1051/0004-6361:20047178}, \href
  {http://adsabs.harvard.edu/abs/2004A%26A...423..537L} {423, 537}

\bibitem[\protect\citeauthoryear{Li, Zhao  \& Li}{Li et~al.}{2007}]{Li+2007}
Li M.,  Zhao G.,   Li A.,  2007, Monthly Notices of the Royal Astronomical
  Society: Letters, 382, L26

\bibitem[\protect\citeauthoryear{Li, Mari{\~n}as  \& Telesco}{Li
  et~al.}{2014}]{Li+2014}
Li D.,  Mari{\~n}as N.,   Telesco C.~M.,  2014, The Astrophysical Journal, 796,
  74

\bibitem[\protect\citeauthoryear{{Li}, {Telesco}, {Zhang}, {Wright}, {Pantin},
  {Barnes}  \& {Packham}}{{Li} et~al.}{2018}]{LI+2018}
{Li} D.,  {Telesco} C.~M.,  {Zhang} H.,  {Wright} C.~M.,  {Pantin} E.,
  {Barnes} P.~J.,   {Packham} C.,  2018, \mn@doi [Monthly Notices of the Royal
  Astronomical Society] {10.1093/mnras/stx2228}, \href
  {http://adsabs.harvard.edu/abs/2018MNRAS.473.1427L} {473, 1427}

\bibitem[\protect\citeauthoryear{{Meeus}, {Waters}, {Bouwman}, {van den
  Ancker}, {Waelkens}  \& {Malfait}}{{Meeus} et~al.}{2001}]{Meeus+2001}
{Meeus} G.,  {Waters} L.~B.~F.~M.,  {Bouwman} J.,  {van den Ancker} M.~E.,
  {Waelkens} C.,   {Malfait} K.,  2001, \mn@doi [Astronomy and Astrophysics]
  {10.1051/0004-6361:20000144}, \href
  {http://adsabs.harvard.edu/abs/2001A%26A...365..476M} {365, 476}

\bibitem[\protect\citeauthoryear{{Meeus} et~al.,}{{Meeus}
  et~al.}{2009}]{Meeus+2009}
{Meeus} G.,  et~al., 2009, \mn@doi [Astronomy and Astrophysics]
  {10.1051/0004-6361/200811490}, \href
  {http://adsabs.harvard.edu/abs/2009A%26A...497..379M} {497, 379}

\bibitem[\protect\citeauthoryear{{Men'shchikov}, {Henning}  \&
  {Fischer}}{{Men'shchikov} et~al.}{1999}]{Menshchikov+1999}
{Men'shchikov} A.~B.,  {Henning} T.,   {Fischer} O.,  1999, \mn@doi [The
  Astrophysical Journal] {10.1086/307333}, \href
  {http://adsabs.harvard.edu/abs/1999ApJ...519..257M} {519, 257}

\bibitem[\protect\citeauthoryear{Messenger, Joswiak, Ito, Matrajt  \&
  Brownlee}{Messenger et~al.}{2009}]{Messenger+2009}
Messenger S.,  Joswiak D.,  Ito M.,  Matrajt G.,   Brownlee D.,  2009, in Lunar
  and Planetary Science Conference.

\bibitem[\protect\citeauthoryear{{Min}, {Waters}, {de Koter}, {Hovenier},
  {Keller}  \& {Markwick-Kemper}}{{Min} et~al.}{2007}]{Min+2007}
{Min} M.,  {Waters} L.~B.~F.~M.,  {de Koter} A.,  {Hovenier} J.~W.,  {Keller}
  L.~P.,   {Markwick-Kemper} F.,  2007, \mn@doi [Astronomy and Astrophysics]
  {10.1051/0004-6361:20065436}, \href
  {http://adsabs.harvard.edu/abs/2007A%26A...462..667M} {462, 667}

\bibitem[\protect\citeauthoryear{{Mishchenko}, {Travis}  \&
  {Lacis}}{{Mishchenko} et~al.}{2002}]{Mishchenko+2002}
{Mishchenko} M.~I.,  {Travis} L.~D.,   {Lacis} A.~A.,  2002, {Scattering,
  absorption, and emission of light by small particles}

\bibitem[\protect\citeauthoryear{{Molster}, {Waters}  \& {Kemper}}{{Molster}
  et~al.}{2010}]{Molster+2010}
{Molster} F.~J.,  {Waters} L.~B.~F.~M.,   {Kemper} F.,  2010, in {Henning} T.,
  ed.,  Lecture Notes in Physics, Berlin Springer Verlag Vol. 815, Lecture
  Notes in Physics, Berlin Springer Verlag. pp 143--201,
  \mn@doi{10.1007/978-3-642-13259-9_3}

\bibitem[\protect\citeauthoryear{{Olofsson} et~al.,}{{Olofsson}
  et~al.}{2009}]{Olofson+2009}
{Olofsson} J.,  et~al., 2009, \mn@doi [Astronomy and Astrophysics]
  {10.1051/0004-6361/200912062}, \href
  {http://adsabs.harvard.edu/abs/2009A%26A...507..327O} {507, 327}

\bibitem[\protect\citeauthoryear{{Orofino}, {Blanco}  \& {Fonti}}{{Orofino}
  et~al.}{1994}]{Orofino+1994}
{Orofino} V.,  {Blanco} A.,   {Fonti} S.,  1994, Astronomy and Astrophysics,
  \href {http://adsabs.harvard.edu/abs/1994A%26A...282..657O} {282, 657}

\bibitem[\protect\citeauthoryear{{Ossenkopf}}{{Ossenkopf}}{1991}]{Ossenkopf+1991}
{Ossenkopf} V.,  1991, Astronomy and Astrophysics, \href
  {http://adsabs.harvard.edu/abs/1991A%26A...251..210O} {251, 210}

\bibitem[\protect\citeauthoryear{{Packham}, {Hough}  \& {Telesco}}{{Packham}
  et~al.}{2005}]{Packham+2005}
{Packham} C.,  {Hough} J.~H.,   {Telesco} C.~M.,  2005, in {Adamson} A.,
  {Aspin} C.,  {Davis} C.,   {Fujiyoshi} T.,  eds,  Astronomical Society of the
  Pacific Conference Series Vol. 343, Astronomical Polarimetry: Current Status
  and Future Directions. p.~38

\bibitem[\protect\citeauthoryear{P{\'e}gouri{\'e} et~al.}{P{\'e}gouri{\'e}
  et~al.}{1988}]{Pegourie+1988}
P{\'e}gouri{\'e} B.,  et~al., 1988, Astronomy and Astrophysics, 194, 335

\bibitem[\protect\citeauthoryear{{Perrin} \& {Sivan}}{{Perrin} \&
  {Sivan}}{1990}]{Perrin+1990}
{Perrin} J.-M.,  {Sivan} J.-P.,  1990, Astronomy and Astrophysics, \href
  {http://adsabs.harvard.edu/abs/1990A%26A...228..238P} {228, 238}

\bibitem[\protect\citeauthoryear{Poppe}{Poppe}{2003}]{Poppe+2003}
Poppe T.,  2003, Icarus, 164, 139

\bibitem[\protect\citeauthoryear{{Poteet} et~al.,}{{Poteet}
  et~al.}{2011}]{Poteet+2011}
{Poteet} C.~A.,  et~al., 2011, \mn@doi [The Astrophysical Journal Letters]
  {10.1088/2041-8205/733/2/L32}, \href
  {http://adsabs.harvard.edu/abs/2011ApJ...733L..32P} {733, L32}

\bibitem[\protect\citeauthoryear{Pott, Eckart, Glindemann, Sch{\"o}del,
  Viehmann  \& Robberto}{Pott et~al.}{2008}]{Pott+2008}
Pott J.-U.,  Eckart A.,  Glindemann A.,  Sch{\"o}del R.,  Viehmann T.,
  Robberto M.,  2008, Astronomy and Astrophysics, 480, 115

\bibitem[\protect\citeauthoryear{{Purcell} \& {Pennypacker}}{{Purcell} \&
  {Pennypacker}}{1973}]{Purcell+1973}
{Purcell} E.~M.,  {Pennypacker} C.~R.,  1973, \mn@doi [The Astrophysical
  Journal] {10.1086/152538}, \href
  {http://adsabs.harvard.edu/abs/1973ApJ...186..705P} {186, 705}

\bibitem[\protect\citeauthoryear{{Saija}, {Iat{\`i}}, {Borghese}, {Denti},
  {Aiello}  \& {Cecchi-Pestellini}}{{Saija} et~al.}{2001}]{Saija+2001}
{Saija} R.,  {Iat{\`i}} M.~A.,  {Borghese} F.,  {Denti} P.,  {Aiello} S.,
  {Cecchi-Pestellini} C.,  2001, \mn@doi [The Astrophysical Journal]
  {10.1086/322350}, \href {http://adsabs.harvard.edu/abs/2001ApJ...559..993S}
  {559, 993}

\bibitem[\protect\citeauthoryear{{Sakon}, {Onaka}, {Okamoto}, {Kataza},
  {Kaneda}  \& {Honda}}{{Sakon} et~al.}{2007}]{Sakon+2007}
{Sakon} I.,  {Onaka} T.,  {Okamoto} Y.~K.,  {Kataza} H.,  {Kaneda} H.,
  {Honda} M.,  2007, in {Ip} W.-H.,  {Bhardwaj} A.,  eds,  Vol. 7, Advances in
  Geosciences. Volume 7: Planetary Science (PS). Edited by Ip, Wing-Huen;
  Bhardwaj, Anil. Published by World Scientific. ISBN \#978-981-2708-92-2,
  2007, pp.143-154. pp 143--154 (\mn@eprint {} {astro-ph/0701215}),
  \mn@doi{10.1142/9789812708922_0014}

\bibitem[\protect\citeauthoryear{{Sicilia-Aguilar}, {Hartmann}, {Watson},
  {Bohac}, {Henning}  \& {Bouwman}}{{Sicilia-Aguilar}
  et~al.}{2007}]{Sicilia-Aguilar+2007}
{Sicilia-Aguilar} A.,  {Hartmann} L.~W.,  {Watson} D.,  {Bohac} C.,  {Henning}
  T.,   {Bouwman} J.,  2007, \mn@doi [The Astrophysical Journal]
  {10.1086/512121}, \href {http://adsabs.harvard.edu/abs/2007ApJ...659.1637S}
  {659, 1637}

\bibitem[\protect\citeauthoryear{{Speck}, {Barlow}  \& {Skinner}}{{Speck}
  et~al.}{1997}]{Speck+1997}
{Speck} A.~K.,  {Barlow} M.~J.,   {Skinner} C.~J.,  1997, \mn@doi [Monthly
  Notices of the Royal Astronomical Society] {10.1093/mnras/288.2.431}, \href
  {http://adsabs.harvard.edu/abs/1997MNRAS.288..431S} {288, 431}

\bibitem[\protect\citeauthoryear{{Suh}}{{Suh}}{2011a}]{Shu+2011}
{Suh} K.-W.,  2011a, \mn@doi [Journal of Korean Astronomical Society]
  {10.5303/JKAS.2011.44.1.013}, \href
  {http://adsabs.harvard.edu/abs/2011JKAS...44...13S} {44, 13}

\bibitem[\protect\citeauthoryear{{Suh}}{{Suh}}{2011b}]{Suh+2011}
{Suh} K.-W.,  2011b, \mn@doi [Journal of Korean Astronomical Society]
  {10.5303/JKAS.2011.44.1.013}, \href
  {http://adsabs.harvard.edu/abs/2011JKAS...44...13S} {44, 13}

\bibitem[\protect\citeauthoryear{{Telesco} et~al.,}{{Telesco}
  et~al.}{2003}]{Telesco+2003}
{Telesco} C.~M.,  et~al., 2003, in {Iye} M.,  {Moorwood} A.~F.~M.,  eds,
  \procspie Vol. 4841, Instrument Design and Performance for Optical/Infrared
  Ground-based Telescopes. pp 913--922, \mn@doi{10.1117/12.458979}

\bibitem[\protect\citeauthoryear{{Vaidya} \& {Gupta}}{{Vaidya} \&
  {Gupta}}{2009}]{Vaidya+2009}
{Vaidya} D.~B.,  {Gupta} R.,  2009, Journal of Quantitative Spectroscopy \&
  Radiative Transfer, \href {http://adsabs.harvard.edu/abs/2009JQSRT.110.1726V}
  {110, 1726}

\bibitem[\protect\citeauthoryear{{Voshchinnikov}, {Il'in}  \&
  {Henning}}{{Voshchinnikov} et~al.}{2005}]{Voshchinnikov+2005}
{Voshchinnikov} N.~V.,  {Il'in} V.~B.,   {Henning} T.,  2005, \mn@doi
  [Astronomy and Astrophysics] {10.1051/0004-6361:200400081}, \href
  {http://adsabs.harvard.edu/abs/2005A%26A...429..371V} {429, 371}

\bibitem[\protect\citeauthoryear{{Voshchinnikov}, {Il'in}, {Henning}  \&
  {Dubkova}}{{Voshchinnikov} et~al.}{2006}]{Voshchinnikov+2006}
{Voshchinnikov} N.~V.,  {Il'in} V.~B.,  {Henning} T.,   {Dubkova} D.~N.,  2006,
  \mn@doi [Astronomy and Astrophysics] {10.1051/0004-6361:20053371}, \href
  {http://adsabs.harvard.edu/abs/2006A%26A...445..167V} {445, 167}

\bibitem[\protect\citeauthoryear{{Waters} \& {Waelkens}}{{Waters} \&
  {Waelkens}}{1998}]{Waters+Waelkens1998}
{Waters} L.~B.~F.~M.,  {Waelkens} C.,  1998, \mn@doi [Annual Review of
  Astronomy and Astrophysics] {10.1146/annurev.astro.36.1.233}, \href
  {http://adsabs.harvard.edu/abs/1998ARA%26A..36..233W} {36, 233}

\bibitem[\protect\citeauthoryear{{Watson} et~al.,}{{Watson}
  et~al.}{2009}]{Watson+2009}
{Watson} D.~M.,  et~al., 2009, \mn@doi [The Astrophysical Journal Supplement]
  {10.1088/0067-0049/180/1/84}, \href
  {http://adsabs.harvard.edu/abs/2009ApJS..180...84W} {180, 84}

\bibitem[\protect\citeauthoryear{Weidenschilling}{Weidenschilling}{2000}]{Weidenschilling+2000}
Weidenschilling S.~J.,  2000, Space Science Reviews, 92, 295

\bibitem[\protect\citeauthoryear{{Whittet}}{{Whittet}}{2003}]{Whittet+2003}
{Whittet} D.~C.~B.,  ed. 2003, {Dust in the galactic environment}

\bibitem[\protect\citeauthoryear{{Wolff}, {Clayton}, {Martin}  \&
  {Schulte-Ladbeck}}{{Wolff} et~al.}{1994}]{Wolff+1994}
{Wolff} M.~J.,  {Clayton} G.~C.,  {Martin} P.~G.,   {Schulte-Ladbeck} R.~E.,
  1994, \mn@doi [The Astrophysical Journal] {10.1086/173817}, \href
  {http://adsabs.harvard.edu/abs/1994ApJ...423..412W} {423, 412}

\bibitem[\protect\citeauthoryear{{Wright}, {Do Duy}  \& {Lawson}}{{Wright}
  et~al.}{2016}]{Wright+2016}
{Wright} C.~M.,  {Do Duy} T.,   {Lawson} W.,  2016, \mn@doi [Monthly Notices of
  the Royal Astronomical Society] {10.1093/mnras/stw041}, \href
  {http://adsabs.harvard.edu/abs/2016MNRAS.457.1593W} {457, 1593}

\bibitem[\protect\citeauthoryear{Zhang et~al.,}{Zhang
  et~al.}{2017a}]{Zhang+2017b}
Zhang H.,  et~al., 2017a, \mn@doi [Monthly Notices of the Royal Astronomical
  Society] {10.1093/mnras/stw2761}, 465, 2983

\bibitem[\protect\citeauthoryear{{Zhang}, {Telesco}, {Hoang}, {Li}, {Pantin},
  {Wright}, {Li}  \& {Barnes}}{{Zhang} et~al.}{2017b}]{Zhang+2017a}
{Zhang} H.,  {Telesco} C.~M.,  {Hoang} T.,  {Li} A.,  {Pantin} E.,  {Wright}
  C.~M.,  {Li} D.,   {Barnes} P.,  2017b, \mn@doi [The Astrophysical Journal]
  {10.3847/1538-4357/aa77ff}, \href
  {http://adsabs.harvard.edu/abs/2017ApJ...844....6Z} {844, 6}

\bibitem[\protect\citeauthoryear{{van Boekel}}{{van
  Boekel}}{2008}]{Boekel+2008}
{van Boekel} R.,  2008, in Journal of Physics Conference Series. p. 012023
  (\mn@eprint {arXiv} {0810.5534}), \mn@doi{10.1088/1742-6596/131/1/012023}

\makeatother
\end{thebibliography}


\bsp	
\label{lastpage}
\end{document}